\input{aipcheck.tex}

  \def\selectedoptions{final}

\documentclass[
   \selectedoptions
  ]
  {aipproc}

  \def\selectedlayoutstyle{8x11single}
\layoutstyle\selectedlayoutstyle

\SetInternalRegister\hbadness{8000} % pseudo latin isn't breaking very well :-)

\newcommand\doingARLO[2][]{%
  \ifx\mmref\undefined #1\else #2\fi
}

\begin{document}

\title{Parsec-scale structure in the warm ISM from polarized galactic radio 
       background observations}


\author{M. Haverkorn}{
  address={Leiden Observatory, P.O. Box 9513, 2300 RA Leiden, the
  Netherlands}, email={haverkrn@strw.leidenuniv.nl} }

\iftrue
\author{P. Katgert}{
  address={Leiden Observatory, P.O. Box 9513, 2300 RA Leiden, the Netherlands},
  email={katgert@strw.leidenuniv.nl}
}

\author{A. G. de Bruyn}{
  address={ASTRON, P.O. Box 2, 7990 AA Dwingeloo, the Netherlands},
  email={ger@astron.nl},
  altaddress={Kapteyn Institute, P.O. Box 800, 9700 AV Groningen, the Netherlands}
}
\fi

\copyrightyear  {2001}

\begin{abstract}
  We present multi-frequency polarization observations of the diffuse
  radio synchrotron background modulated by Faraday rotation, in two
  directions of positive latitude. No extended total intensity $I$ is
  observed, which implies that total intensity has no structure on
  scales smaller than approximately a degree. Polarized intensity and
  polarization angle, however, show abundant small-scale structure on
  scales from arcminutes to degrees. Rotation Measure (RM) maps show
  coherent structure over many synthesized beams, but also abrupt
  large changes over one beam. RM's from polarized extragalactic point
  sources are correlated over the field in each of the two fields,
  indicating a galactic component to the RM, but show no correlation
  with the RM map of the diffuse radiation. The upper limit in
  structure in $I$ puts constraints on the random and regular
  components of the magnetic field in the galactic interstellar medium
  and halo.  The emission is partly depolarized so that the observed
  polarization mostly originates from a nearby part of the
  medium. This explains the lack of correlation between RM from
  diffuse emission and from extragalactic point sources as the latter
  is built up over the entire path length through the medium.
\end{abstract}

\date{\today}

\maketitle

\section{Introduction}

Synchrotron radiation emitted in our galaxy provides a diffuse radio
background, which is altered by Faraday rotation and depolarization
when it propagates through the galactic halo and interstellar medium.
Multi-frequency observations of the linearly polarized component of
this radio background allow determination of the Rotation Measure (RM)
of the medium along many contiguous lines of sight, from which the
structure of the small-scale galactic magnetic field, weighted with
electron density, can be probed. After the discovery paper
(\cite{wieringa}), many other high resolution radio background
observations have shown intriguing polarization structure, mostly in
the galactic plane (\cite{uyaniker}, \cite{gray},
\cite{gaensler}, \cite{duncan}). Here, we focus  on regions of 
positive galactic latitude to evade complexities such as HII regions,
supernova remnants etc. Furthermore, these are sensitive and
multi-frequency observations so that accurate RM values can be
determined.

\section{Observations}

With the Westerbork Synthesis Radio Telescope (WSRT) we mapped the
polarized radio background in two fields of over 50 square degrees
each in the second galactic quadrant at positive latitudes. All four
Stokes parameters $I$, $Q$, $U$, and $V$ were imaged in five frequency
bands centered at 341, 349, 355, 360, and 375~MHz simultaneously
(bandwidth 5 MHz), at a resolution of $\sim$4$^{\prime}$. The first
field, in the constellation Auriga, is 7$^{\circ}\times 9^{\circ}$ in
size and centered on $(l,b) = (161, 16)^{\circ}$; the second field, in
the constellation Horologium, is 8$^{\circ}\times 8^{\circ}$ wide and
centered on $(l,b) = (137, 7)^{\circ}$.  No total intensity $I$ was
detected (besides point sources) in either field down to $\sim$0.7~K,
which is $< 1.5\%$ of the expected sky brightness in these regions,
indicating that $I$ does not vary on scales detectable to the
interferometer, i.e.\ below about a degree.  However, linearly
polarized intensity $P$ and polarization angle $\varphi$ show abundant
small-scale structure. Other fields observed with the WSRT at a single
frequency around 350 MHz also show small-scale structure in polarized
intensity and polarization angle, but of very different topologies
(\cite{katgert}).

\section{Analysis of the Auriga field}

%================================
\begin{figure}
  \includegraphics[height=0.44\textheight]{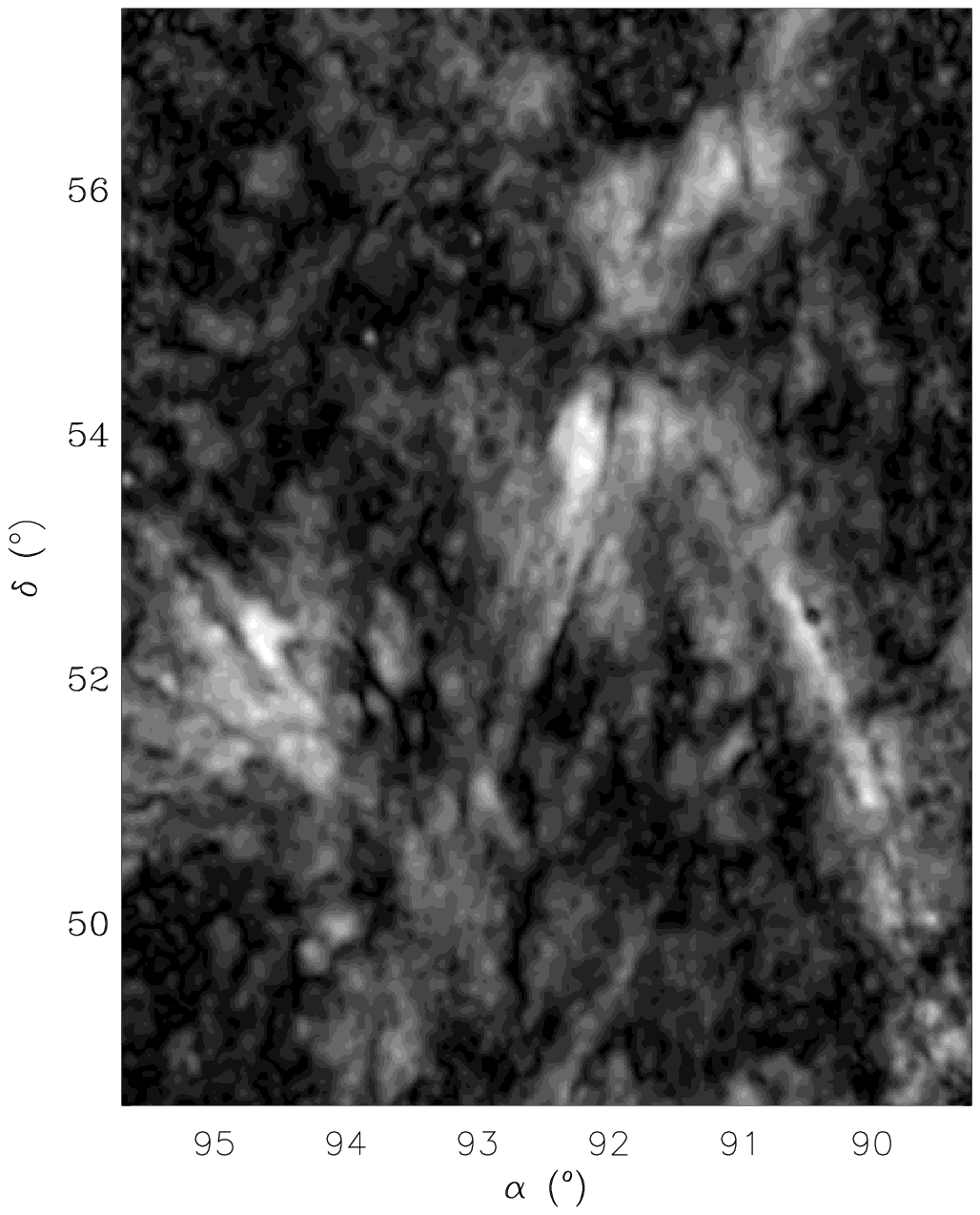}
  \includegraphics[height=0.44\textheight]{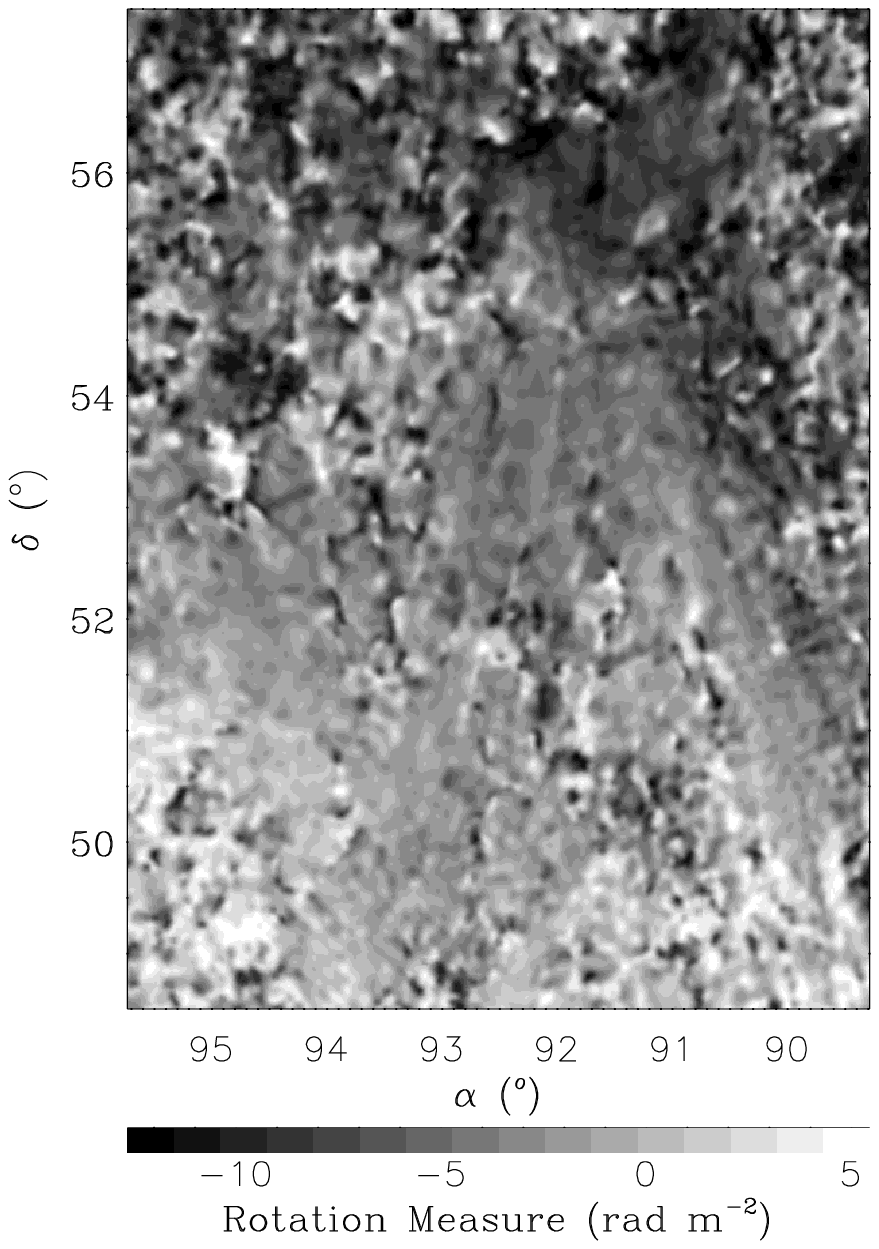} 
  \caption{{\it Left:} polarized intensity $P$ at 349~MHz in the Auriga 
  field at 4$^{\prime}$ resolution. White denotes a maximum $T_{b,pol}
  \approx 18$~K. {\it Right:} RM in the Auriga field. Very high or low
  RM values ($|\mbox{RM}| \approx 30 - 60$ rad m$^{-2}$) in the field
  have been removed from the maps (see text). } 
  \label{fig_aur}
\end{figure}
%================================

The structure in polarized intensity $P$ in the Auriga field shows a
wide variety in topology on several scales, as shown in the left plot
of Fig.~\ref{fig_aur}, where $P$ at 349 MHz is mapped at 4$^{\prime}$
resolution.  The typical polarization brightness temperature is
$T_{b,pol} \approx$~6~-~8~K, with a maximum of $\sim$18~K.  From the
Haslam continuum survey at 408 MHz (\cite{haslam}), the $I$-background
at 408 MHz in this region of the sky is $\sim$33~K.  Extrapolating
this to our frequencies with a temperature spectral index of $-2.5$
between 341~MHz and 408~MHz (\cite{roger}), the total intensity
background is $\sim$41~-~52~K. So the maximum degree of polarization
$p_{max} \approx$ 35\%, with an average $p$ of 15\%.

In addition, a pattern of black narrow wiggly canals is visible (see
e.g.\ the canal around ($\alpha,\delta$) = (92.7, 49 - 51)$^{\circ}$).
These canals are all one synthesized beam wide and have been shown to
separate regions of fairly constant polarization angle $\varphi$ where
the difference in $\varphi$ is approximately 90$^{\circ}$ ($\pm\, n\,
180^{\circ}$, $n=1,2,3\ldots$), which causes beam depolarization
(\cite{haverkorn}). The angle changes are due to abrupt changes in RM.
Hence, the canals reflect specific features in the angle distribution.
Other angle (and RM) changes within the beam cause less or no
depolarization, so that they do not leave easily visible traces in the
polarized intensity distribution.

\begin{figure}
  \includegraphics[height=.24\textheight]{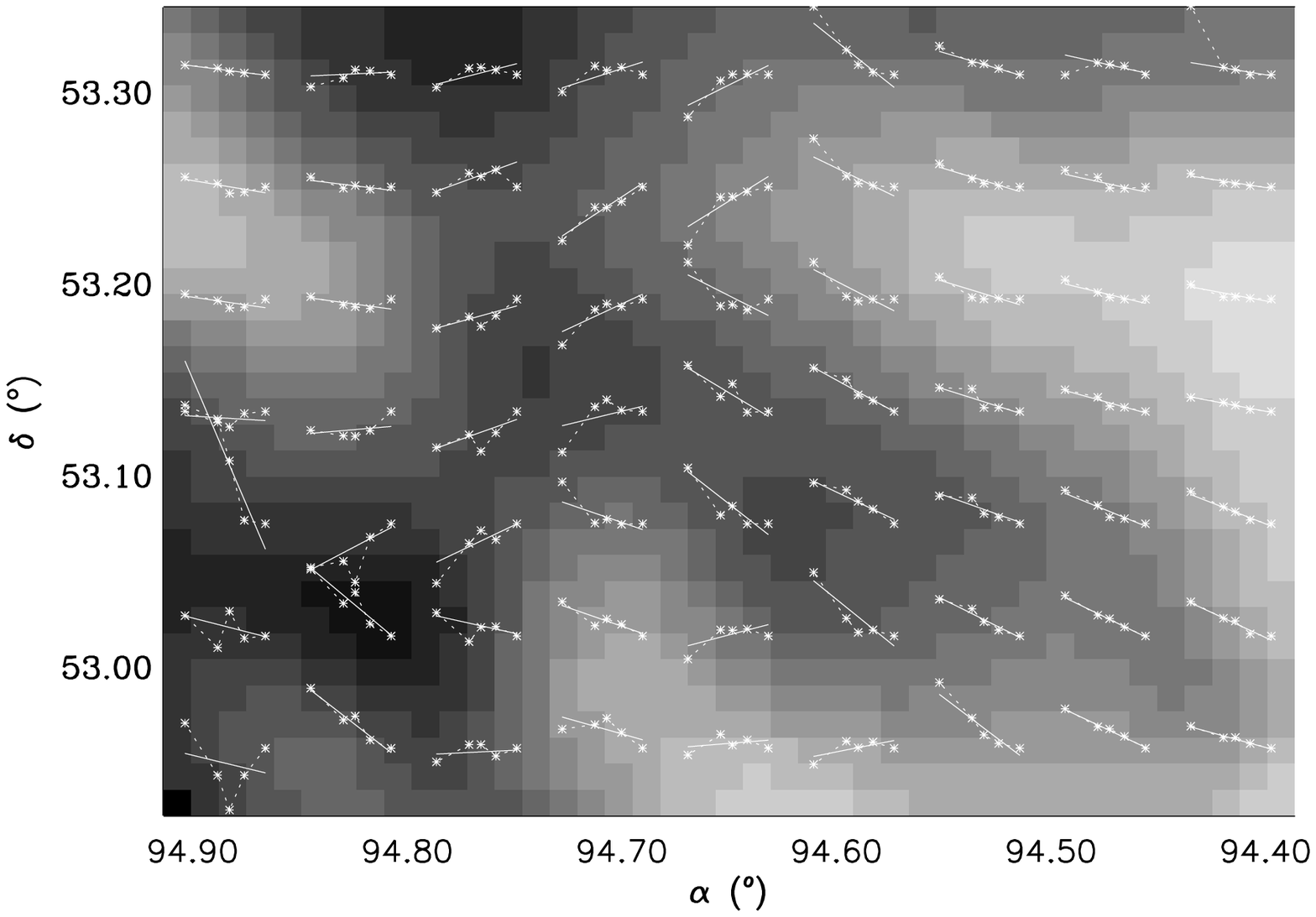}
  \includegraphics[height=.24\textheight]{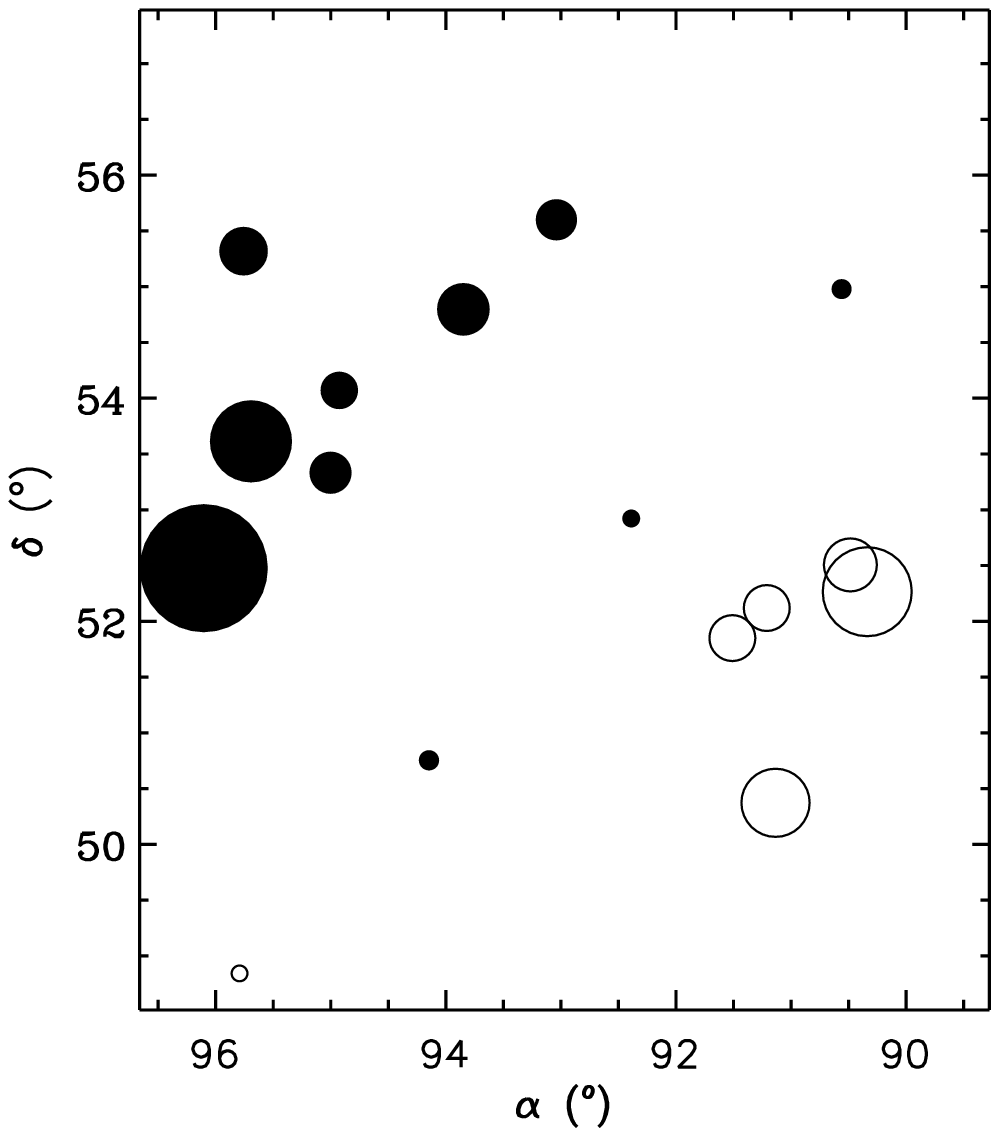}
  \caption{{\it Left:} Graphs of polarization angle against the
           square of the wavelength, so that the slope is RM. Each
           graph denotes an independent beam. The greyscale denotes
           polarized intensity at 349 MHz, five times oversampled.
           RM's range from $\sim$$-10$ to 10 rad m$^{-2}$, ignoring
           the one anomalously large negative RM on the left
           side. Abrupt changes in Rotation Measure over one beam
           (4$^{\prime}$) are coherent along several beams. {\it
           Right:} RM's of observed polarized extragalactic point
           sources in the Auriga field. The radii of the circles are
           scaled with magnitude of RM, where filled circles are
           positive RM's.  Maximum (minimum) RM is 19.5 ($-13.6$)
           rad~m$^{-2}$.}
  \label{fig_largedrm}
\end{figure}

The RM of the Faraday-rotating material can be derived from
$\varphi(\lambda^2) \propto \mbox{RM}\,\lambda^2$ (see
\cite{haverkorn2} for details and pitfalls).  The right plot in
Fig.~\ref{fig_aur} gives a 4$^{\prime}$ resolution RM map of the
Auriga field. The average RM $\approx -3.4$ rad~m$^{-2}$, and in
general $|\mbox{RM}| < 15$ rad m$^{-2}$. Very high or low RM values
($|\mbox{RM}| \approx 30 - 60$ rad m$^{-2}$) in the field occur only
at positions where polarized intensity $P$ is very low, so noise
errors in polarization angle are very large. Therefore the RM's at
these positions are not reliable and have been removed from the
maps. RM's show structure on scales of many beams (up to degree
scales), but also abrupt changes from one beam to another. This is
illustrated in the left map of Fig.~\ref{fig_largedrm}, where a small
part of the Auriga field is shown. Here, the lines are graphs of
polarization angle against wavelength squared, so that the slope of
the line is RM. Each graph is an independent synthesized beam. The
greyscale denotes polarized intensity $P$ at 349 MHz, five times
oversampled. Large sudden RM changes occur: e.g. at ($\alpha,\delta$)
= (94.68,53.15)$^{\circ}$, RM changes from $-9$ rad m$^{-2}$ to 7 rad
m$^{-2}$, and at ($\alpha,\delta$) = (94.60,53.00)$^{\circ}$ from 3
rad m$^{-2}$ to $-11$ rad m$^{-2}$. A change in {\it sign} of RM
indicates in general a change in {\it direction} of the galactic
magnetic field along the line of sight, although numerical models of
propagation of polarized radiation through a mixed (synchrotron
emitting and Faraday-rotating) medium show a change of sign of RM also
without a reversal in the magnetic field (\cite{sokoloff}).

%=============================
%==============================

We detected seventeen polarized extragalactic sources in the Auriga
field at a higher resolution ($\sim$1$^{\prime}$), with RM's from
$-13.6$ to 19.5 rad~m$^{-2}$.  The right plot in
Fig.~\ref{fig_largedrm} shows the RM's and positions of the sources,
where the sizes of the circles are proportional to RM, and open
(filled) circles denote negative (positive) RM's. The RM's of the
extragalactic sources exhibit a clear gradient across the field of
$\sim$5 rad m$^{-2}$ per degree roughly in the direction of galactic
latitude, indicating a galactic component to the RM's of the
sources. The change of sign over the field means a (local) reversal in
the magnetic field parallel to the line of sight. We estimate a RM
component intrinsic to the source of $<$ 5~rad~m$^{-2}$, consistent
with earlier estimates (\cite{leahy}).  The RM structure of the
diffuse galactic radiation is independent from the observed RM of the
extragalactic sources (see below).

\section{Analysis of the Horologium field}

%================================
\begin{figure}
  \includegraphics[height=.36\textheight]{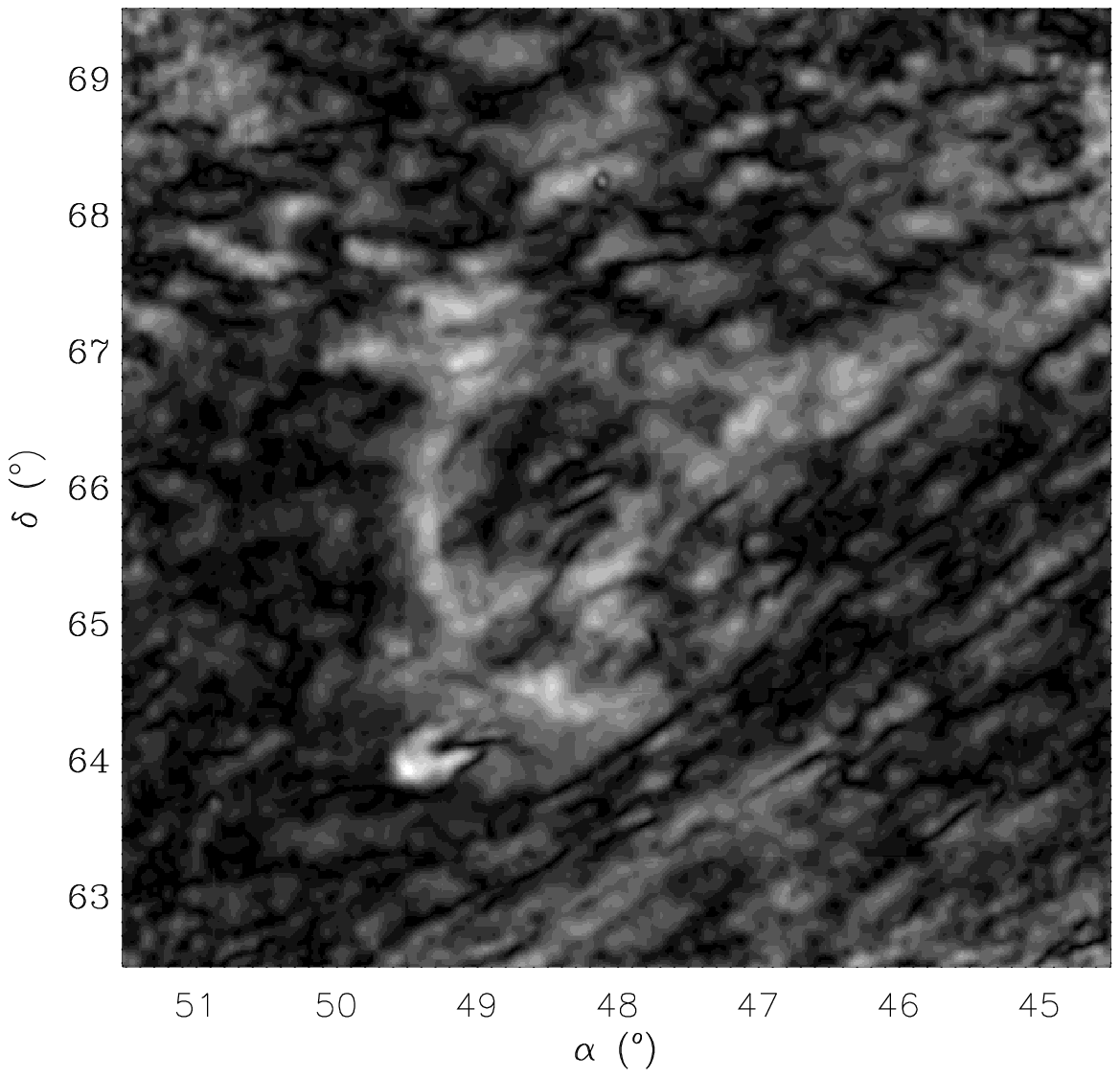}
  \includegraphics[height=.36\textheight]{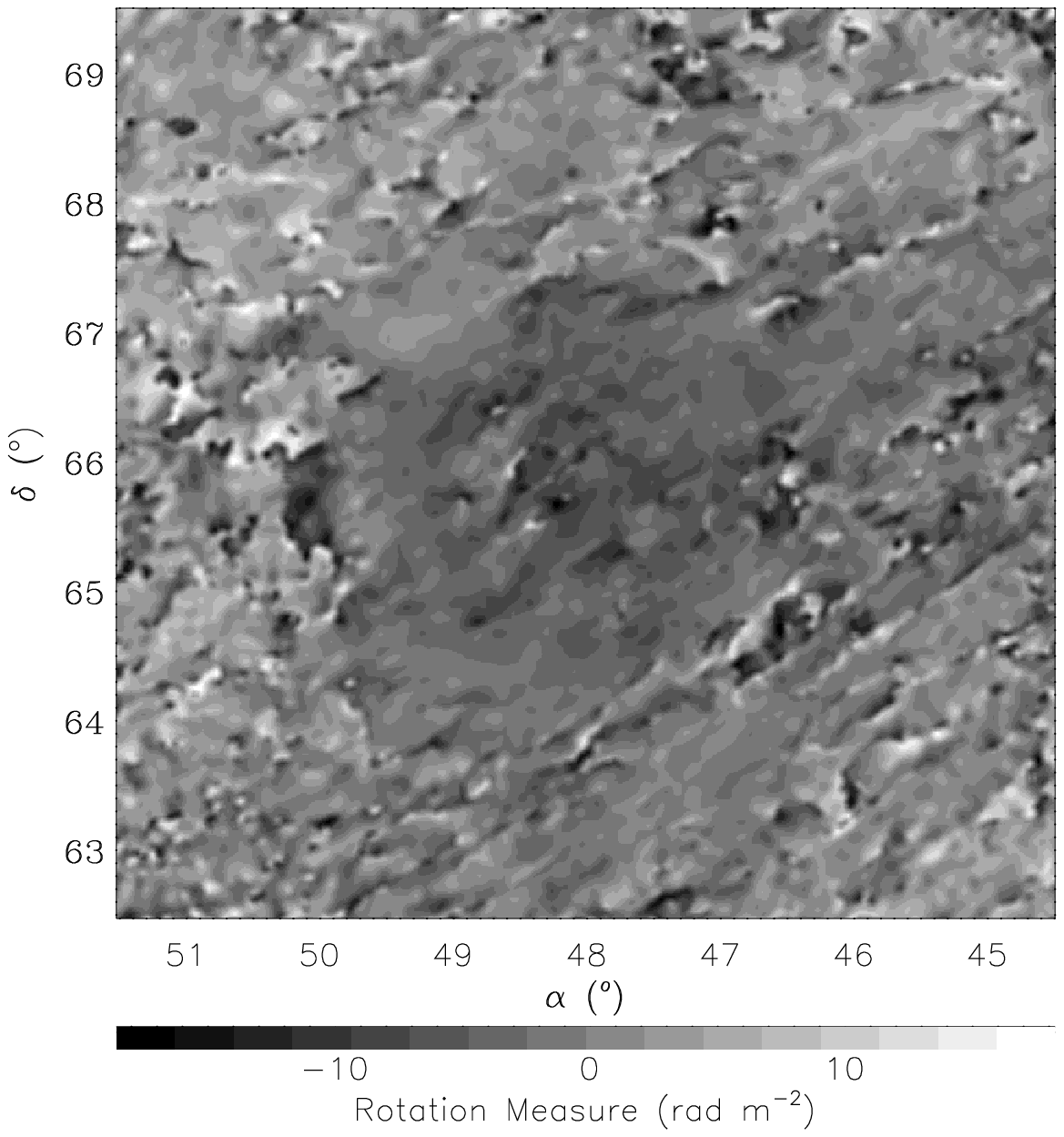}
  \caption{{\it Left:} polarized intensity $P$ at 349~MHz in the Horologium
           field at 4$^{\prime}$ resolution. White denotes a maximum
           $T_{b,pol} \approx$~15~K. {\it Right:} RM in the
           Horologium field. Very high or low RM values ($|\mbox{RM}|
           \approx 30 - 60$ rad m$^{-2}$) in the field have been
           removed from the maps (see text).}  
  \label{fig_hor}
\end{figure}
%================================

The left map in Fig.~\ref{fig_hor} shows polarized intensity at
349~MHz in the Horologium field. The average polarization brightness
temperature is $\sim$5 K, and $T_{b,pol}$ in the ring-like structure
is $\sim$ 11-15 K. The average degree of polarization is 5\% and the
maximum 25\%, again derived using the Haslam survey
(\cite{haslam}). The ring with diameter $\sim$2.7$^{\circ}$ is visible
in all frequency bands, although the ring becomes more diffuse and
smeared out towards higher frequencies and the left side is clearer
than the right side. Beam-wide depolarized canals are again caused by
beam depolarization. In the lower right, the depolarized canals are
aligned along constant latitude.  Caution is required in interpreting
the $P$ map, as the Horologium field is imbedded in a region of very
high constant polarization (\cite{brouw}). Due to missing small
spacings, we cannot detect this large-scale component in $Q$ and/or
$U$, which causes a distorted image in $P$ as well as RM. Possible
solutions to this problem will be given in a forthcoming paper.  A
ring in $P$ was also detected by Verschuur (\cite{verschuur}) at
40${^\prime}$ resolution with a single dish, although not as an
enhancement but as a deficiency in $P$.

\begin{figure}
  \includegraphics[height=.17\textheight]{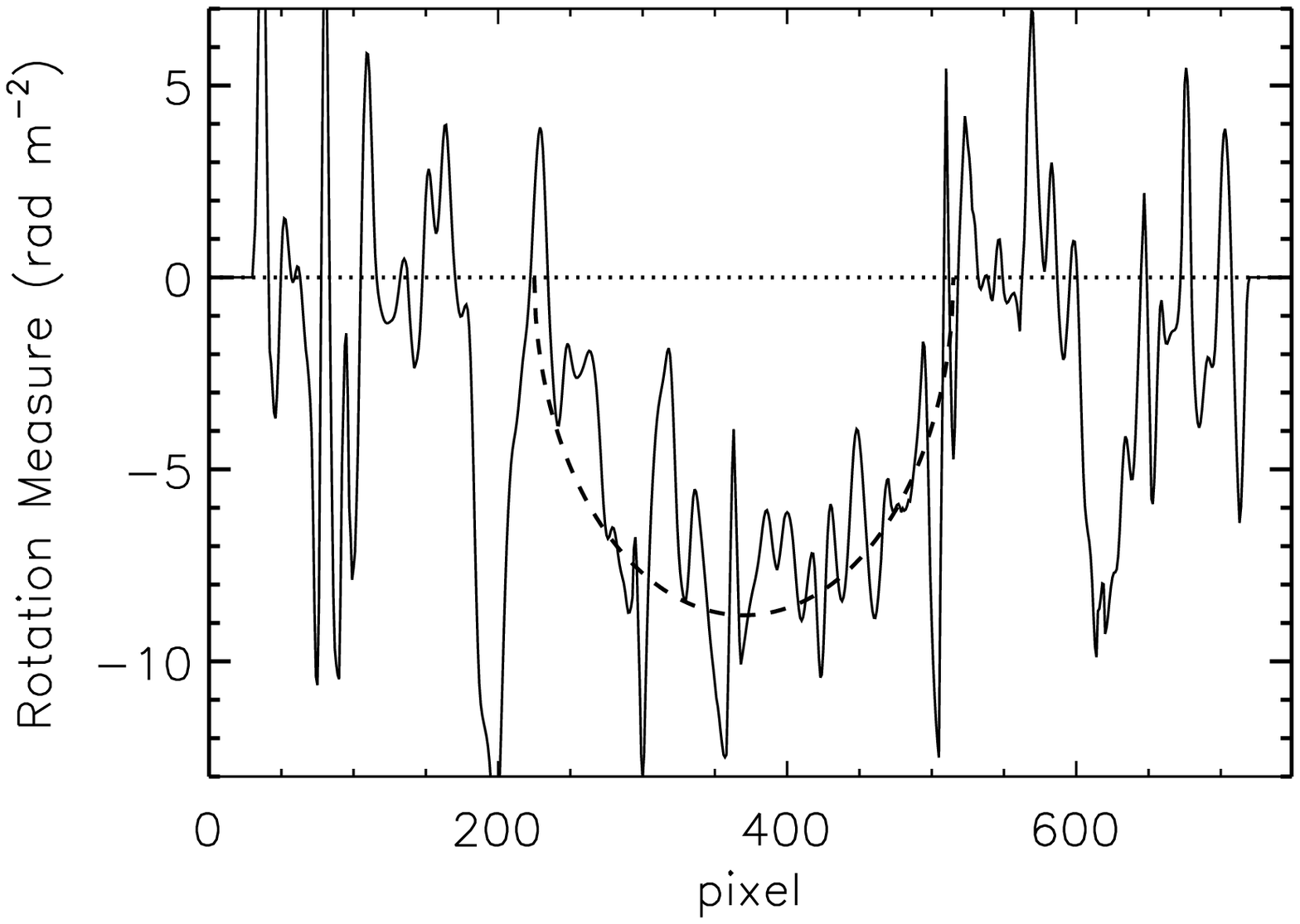}
  \includegraphics[height=.17\textheight]{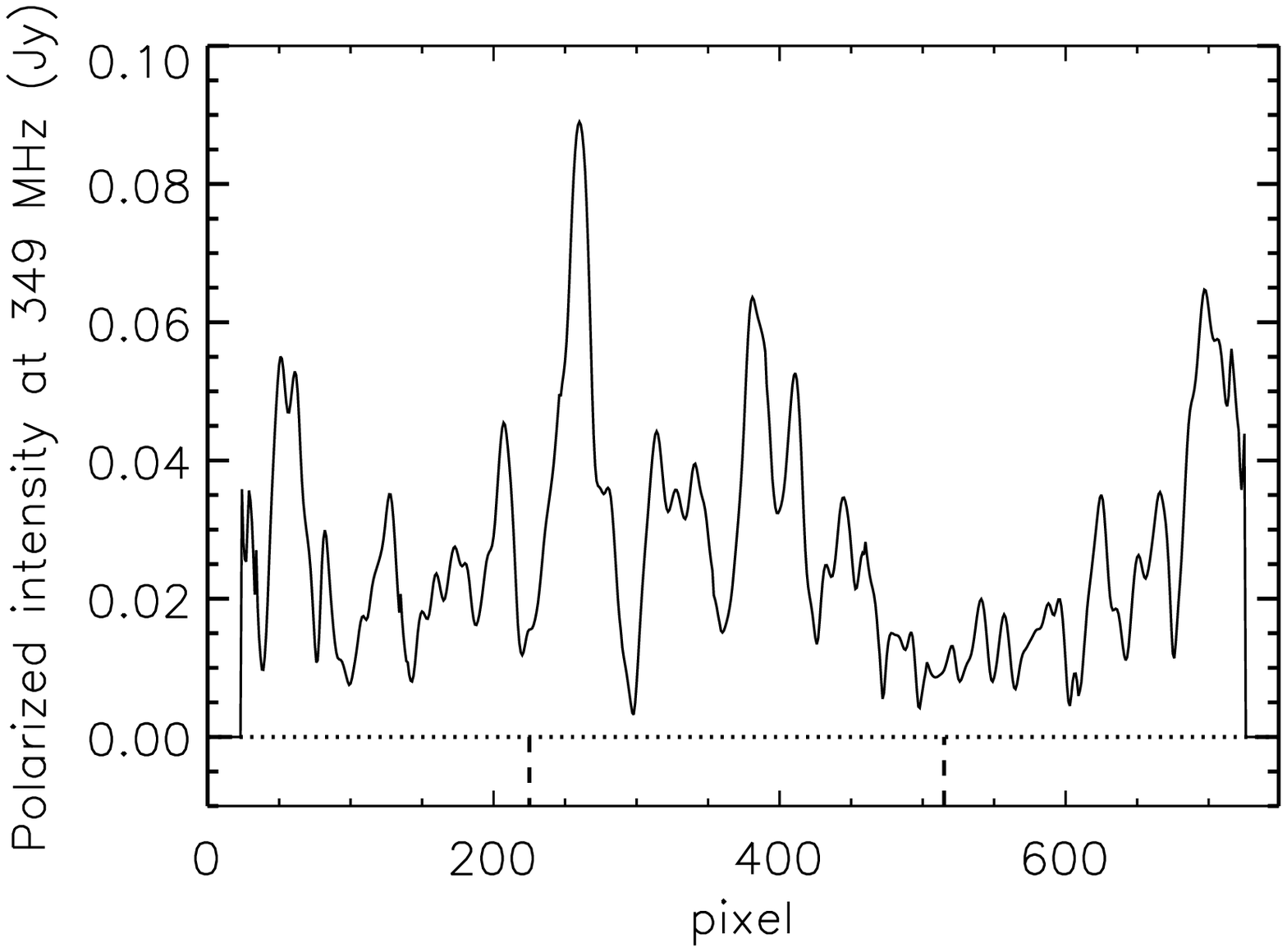}
  \includegraphics[height=.21\textheight]{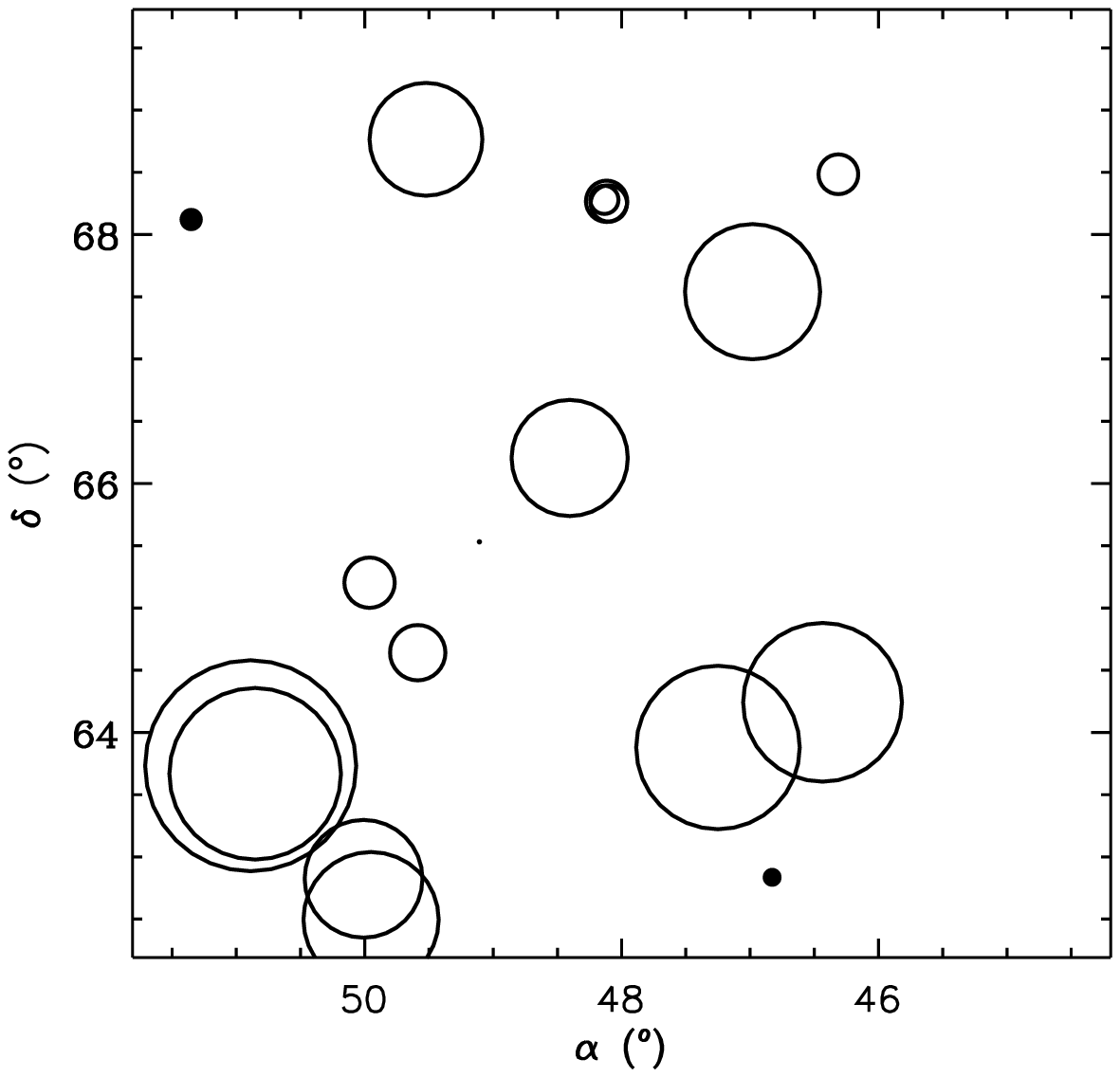}
  \caption{{\it Left and center:} horizontal cross sections through
  the RM map and the $P$ map of the Horologium field in
  Fig.~\ref{fig_hor} (at 4$^{\prime}$ resolution) through the center
  of the RM disk. The dashed line in the RM plot is a fit to the
  decrease in RM, calculated for a sphere of constant thermal electron
  density and line-of-sight magnetic field. In the $P$ plot, the same
  fit is indicated at the bottom. {\it Right:} RM's of observed
  polarized extragalactic point sources in the Horologium field. The
  radii of the circles are again scaled with magnitude of RM, but with
  radii twice as small as in Fig.~\ref{fig_largedrm}.  Maximum
  (minimum) RM is 7.4 ($-67.9$) rad m$^{-2}$.}  \label{fig_crossring}
\end{figure}

In the RM map of the Horologium field (the right hand map of
Fig.~\ref{fig_hor}), a circular structure is clearly visible, which is
slightly bigger than the ring in polarized intensity. Inside this RM
disk, RM's decrease from the edge of the ring to the center, from
$\sim$0 to $-10$ rad m$^{-2}$. Outside the disk, RM's vary around zero
without a clear gradient, with a maximum of $\sim$7 rad m$^{-2}$. Note
that if $P$ is low outside the disk, the influence of undetected
large-scale polarization becomes larger and RM's may not be
well-determined. The left and center plots of Fig.~\ref{fig_crossring}
give horizontal cross sections through the center of the RM disk and
through the $P$ ring at 349 MHz at the same position. The decrease in
RM within the RM disk can be modeled with a homogeneous sphere of
constant electron density and magnetic field. For values of electron
density $n_e = 0.03$ cm$^{-3}$ and parallel magnetic field
$B_{\parallel} = 2\, \mu$G, the distance to the sphere is about 300 pc.

In the Horologium field, we detected 18 polarized extragalactic point
sources as shown in the right plot of Fig.~\ref{fig_crossring}. RM's
of these sources are in the range of $-67.9$ to 7.4 rad
m$^{-2}$. These RM values are not correlated with the RM's from the
diffuse radiation, similar as in the Auriga field. Extragalactic
source RM's are higher than in the Auriga field, which is most likely
due to the lower latitude of the Horologium field.

%==============================
%===============================

\section{Interpretation of the observations}

The galactic synchrotron emission is thought to originate from two
separate domains centered on the galactic plane: a thin and a thick
disk, with scale heights of 180 pc and 1.8 kpc respectively
(\cite{beuermann}). The thick disk emits 90\% of the synchrotron
radiation, and the thin disk coincides approximately with the stellar
disk and the HI disk. Pulsar RM observations have shown that the
random magnetic field in the thin disk $B_{ran} \approx 5 \mu$G is of
the same order or larger than the regular magnetic field
(\cite{rand},\cite{ohno}).

The thermal electrons that cause the Faraday rotation of the
synchrotron background are contained in the Reynolds layer, with a
scale height of about a kpc (\cite{reynolds}). The thermal electrons
are more confined to the galaxy than the relativistic electrons, so
there is a ``halo'' above the Reynolds layer that only contains
relativistic but no thermal electrons.

These two mediums thus define three domains in the galactic ISM and
halo with different characteristics, as sketched in
Table~\ref{table}. The thin synchrotron disk (domain I) extends to a
few hundred parsec, and is mixed with the lower parts of the Reynolds
layer and the thick disk. The Local Bubble is not taken into
account. The upper part of the Reynolds layer is also mixed with the
thick synchrotron disk (domain II), whereas the highest part of the
thick synchrotron disk is so high above the galactic plane that it
doesn't contain a significant amount of thermal electrons anymore
(domain III).  Our observations, in particular the very low upper
limit on small-scale structure in total intensity $I$, put strict
constraints on the characteristics of these three domains.

The first constraint is a small scale height of the thin disk. Due to
the large random magnetic field component in the thin disk, the
emissivity $I$ has a fluctuating component. If the observed integrated
emissivity of this fluctuating component is only a few Kelvin, the
$I$-structure can be averaged out if it is on small enough scales,
i.e. if there are enough ``turbulent cells'' along the line of
sight. Therefore only a thin layer is allowed where the $B_{ran}$
component is large, with structure on small enough scales to
average out the small-scale structure in $I$ that is created in this
layer. Typically, in a layer with a scale height of 200 pc,
$B_{ran}$-structure on scales of 10 to 20 pc can smooth $I$ down to
the observed limits.

The second constraint is a constraint on the magnetic field in the
layers above the thin synchrotron disk. Small-scale structure in $I$
emitted in these layers has to be negligible. Assuming equipartition
in energy between the relativistic electrons and magnetic field, the
synchrotron emissivity $\epsilon \propto B^2$. As the fluctuation in
$I$ requires $\Delta\epsilon < 1\%$, this implies that
$B_{ran}/B_{reg} < 0.1$. Therefore, the absence of small-scale
structure in $I$ dictates a regular magnetic field dominating over the
random magnetic field component by more than a factor ten in the halo
of the galaxy, i.e. above a scale height of a few hundred parsecs, at
least for the component of the magnetic field perpendicular to the
line of sight.

The observed structure in polarized intensity is made by
depolarization on small scales in the layers containing thermal
electrons, by three mechanisms.  The first mechanism is beam
depolarization, as discussed above. Second, linearly polarized
radiation emitted at different depths is Faraday-rotated by different
amounts (differential Faraday rotation \cite{gardner}) and third,
small-scale structure in thermal electron density and/or magnetic
field causes spatial structure in Faraday rotation (internal Faraday
dispersion
\cite{burn}). These depolarization mechanisms define a wavelength 
dependent Faraday depth for the polarized radiation, which indicates
that most of the observed polarized intensity comes from the nearer
part of the medium. This explains the difference between RM structure
from the diffuse emission and from extragalactic point sources, as the
latter is built up over the whole path length through the
Faraday-rotating medium (domains I and II). A more quantitative
discussion can be found in Haverkorn et al.\ (\cite{haverkorn2})

\begin{table}
\begin{tabular}{|c|ccc|c|c|l|}
\hline
 {\bf domain} & \multicolumn{3}{|c|}{\bf components} & {\bf scale height (pc)} & 
 {\bf relevant constituents} &  $\mathbf{B_{ran}/B_{reg}}$ \\
\hline
   &    &\multicolumn{1}{c|}{\mbox{}}  &     &    & & \\
III&    &\multicolumn{1}{c|}{\mbox{}}  &Thick&1800&$n_{rel}$, $B$ & $\leq$ 0.1 \\
   &    &\multicolumn{1}{c|}{\mbox{}}  &disk &    & & \\
\cline{1-1} \cline{3-3} \cline{5-7}
   &    &\multicolumn{1}{|c|}{\mbox{} }&     &    & & \\
II &    &\multicolumn{1}{|c|}{Reynolds}&     &1000&$n_{rel}$, $B$, $n_{th}$&$\leq$ 0.1\\
   &    &\multicolumn{1}{|c|}{layer}   &     &    & & \\
\cline{1-2} \cline{5-7}
   &    &\multicolumn{1}{|c|}{\mbox{}} &     &    & & \\
I  &Thin&\multicolumn{1}{|c|}{\mbox{}} &     & 180&$n_{rel}$, $B$, $n_{th}$&$\geq$ 1\\
   &disk&\multicolumn{1}{|c|}{\mbox{}} &     &    & & \\
\hline
\end{tabular}
\caption{Three schematic domains with different characteristics in the 
galactic ISM and halo.}
\label{table}
\end{table}

\section{Conclusions}

We observed two fields of over 50 square degrees, both located in the
second galactic quadrant at positive latitude, in the constellations
Auriga and Horologium. All four Stokes parameters were derived from
observations done at five frequencies 341, 349, 355, 360, and 375 MHz
simultaneously. The total intensity $I$ emission is featureless on
scales smaller than approximately a degree, while linear polarizations
$Q$ and $U$ show abundant structure on arcminute to degree scales.
Polarized intensity has a maximum $T_{b,pol} \approx$ 15 - 18 K.

The observed structure in polarized intensity $P$ is 'cloudy' on
scales from arcminutes up to a degree. Long canals of one synthesized
beam wide where no polarization is detected are caused by beam
depolarization in a beam which separates two regions where the
polarization angle changes abruptly by 90$^{\circ}$ (or 270$^{\circ}$,
540$^{\circ}$ etc).

Values of RM in the two fields are in general small: |RM| $<$ 15 rad
m$^{-2}$. RM maps show coherent structure in RM over several
independent beams up to a degree, but also sudden large RM changes
across one beam of more than 100\%. Not only the magnitude of the RM
changes but regularly also the sign, which is most easily explained by
a change in direction of $B_{\parallel}$. In the Horologium field, a
ring-like structure in polarized intensity coincides with a disk of
radially increasing RM, although the RM disk is slightly larger than
the ring in $P$.

The lack of observed small-scale structure in total intensity $I$ puts
constraints on the medium (the galactic ISM and halo). In the thin
disk of synchrotron radiation, comparable to the stellar and HI disk,
the random magnetic field component $B_{ran}$ is comparable to or
larger than the regular field component $B_{reg}$. As $B_{ran}$ causes
fluctuating $I$, the layer cannot be thicker than a few hundred
parsecs, and the structure has to be on small enough scales to average
out the created $I$ fluctuations. Furthermore, in the thick
synchrotron disk above the thin disk, the random magnetic field has to
be very small: $B_{ran}
\leq 0.1\, B_{reg}$. Because of depolarization in the synchrotron emitting 
and Faraday-rotating medium most of the polarized emission we observe
originates from the nearby medium. As the RM of an extragalactic point
source is built up along the entire path length through the medium,
the RM structure of the diffuse emission can be uncorrelated with the
structure in RM values of extragalactic point sources, as is observed.

\begin{theacknowledgments}
The Westerbork Synthesis Radio Telescope is operated by the
Netherlands Foundation for Research in Astronomy (ASTRON) with
financial support from the Netherlands Organization for Scientific
Research (NWO). This work is supported by NWO grant 614-21-006.
\end{theacknowledgments}

% choose bibtex style depending on layout style and options used in
% sample:

\doingARLO[\bibliographystyle{aipproc}]
          {\ifthenelse{\equal{\AIPcitestyleselect}{num}}
             {\bibliographystyle{arlonum}}
             {\bibliographystyle{arlobib}}
          }
\bibliography{haverkorn}

\end{document}